          [2001/04/25 1.1 (PWD)]
\documentclass[a4paper,twocolumn]{esapub} % European paper
\usepackage{natbib,graphicx}
\title{INTEGRAL monitoring of the Black Hole candidate 1E 1740.7-2942}
\author[1]{\mbox{M. Del Santo}}
\author[1]{\mbox{A. Bazzano}}
\author[2]{\mbox{D. M. Smith}}
\author[3]{\mbox{L. Bassani}}
\author[4]{\mbox{A. J. Bird}}
\author[5]{\mbox{L. Bouchet}}
\author[6]{\mbox{M. Cadolle-Bel}}
\author[1]{\mbox{F. Capitanio}}
\author[1]{\mbox{G. De Cesare}}
\author[6]{\mbox{M. Falanga}}
\author[7]{\mbox{F. Frontera}}
\author[6]{\mbox{P. Goldoni}}
\author[6]{\mbox{A. Goldwurm}}
\author[8]{\mbox{J. Huovelin}}
\author[9]{\mbox{E. Kuulkers}}
\author[10]{\mbox{N. Lund}}
\author[3]{\mbox{G. Malaguti}}
\author[3]{\mbox{A. Malizia}}
\author[6]{\mbox{I. F. Mirabel}}
\author[1]{\mbox{L. Natalucci}}
\author[11]{\mbox{A. Paizis}}
\author[6]{\mbox{J. Paul}}
\author[12]{\mbox{V. Reglero}}
\author[1]{\mbox{P. Ubertini}}
\author[8]{\mbox{O. Vilhu}}
\author[13]{\mbox{A. Zdziarski}}
\author[9]{\mbox{C. Winkler}}
\affil[1]{\it{Istituto di Astrofisica Spaziale e Fisica cosmica/CNR, via del Fosso del Cavaliere 100, 
00133 Roma, Italy}}
\affil[2]{\it{Department of Physics, University of California, Santa Cruz, Santa Cruz, CA 95064}}
\affil[3]{\it{IASF/CNR, sez. Bologna, via Gobetti 101, 40129 Bologna, Italy}}
\affil[4]{\it{School of Physics and Astronomy, University of Southampton, SO17 1BJ, UK}}
\affil[5]{\it{Centre d'$\acute{E}$tude Spatiale des Rayonnements, CNRS/UPS, BP 4346, 31028 Toulouse, France}}
\affil[6]{\it{CEA Saclay, DSM/DAPNIA/SAp, 91191 Gif-sur-Yvette Cedex, France}}
\affil[7]{\it{Dipartimento di Fisica, University of Ferrara, via del Paradiso 12, 44100 Ferrara, Italy}}
\affil[8]{\it{Observatory, PO Box 14, 00014, University of Helsinki, Finland}}
\affil[9]{\it{Research and Scientific Support Department of ESA, ESTEC, Postbus 299, 2200 AG Noordwijk, 
The Netherlands}}
\affil[10]{\it{Danish Space Research Institute, Juliane Maries Vej 30, 2100 Copenhagen, Denmark}}
\affil[11]{\it{ISDC, Chemin d'$\acute{E}$cogia 16, 1290 Versoix, Switzerland}}
\affil[12]{\it{GACE, Universidad de Valencia, PO Box 20085, 46071 Valencia, Spain}}
\affil[13]{\it{Nicolaus Copernicus Astronomical Center, Bartycka 18, 00-716 Warszawa, Poland}}

\def\INTEGRAL{{\it INTEGRAL~}}

\begin{document}


\maketitle

\begin{abstract}
The brightest persistent Galactic black hole candidate close to the Galactic Centre, \mbox{1E 1740.7-2942}, 
has long been observed with {\it{INTEGRAL}}.\\
In this paper, we report on the long-term hard X-ray monitoring obtained during the first 
year of observations as part of the Galactic Centre Deep Exposure.\\
We discuss the temporal and spectral behaviours in different energy bands up to 250 keV, 
as well as the hardness-flux correlations. 
\end{abstract}

\section{Introduction}
The black hole candidate (BHC) \mbox{1E 1740.7-2942} was first observed with 
the {\it{EINSTEIN}} observatory by Hertz \& Grindlay (1984).
It lies $50'$ away from the centre of the Galaxy and it is the brightest persistent source
within a few degrees of the Galactic Centre (GC) above $\sim$40~keV \cite{sunyaev91}.
The source usually shows a hard X-ray spectrum characteristic of BHCs in their low/hard state;
the X-ray emission is strongly attenuated by a high hydrogen column density
$N_{H}=1\times 10^{23}$~cm$^{-2}$ \cite{sakano99}.\\
Interest in this source increased when Mirabel et al. (1992) discovered
double-side jets in radio emission that classified the source as microquasar.
It has also been suggested as a possible positron-electron annihilation source
due to a SIGMA (French instrument on-board of the {\it{GRANAT}} satellite) observation 
of the 511~keV line \cite{bouchet91} from 1E 1740.7-2942.
Actually, a definitive confirmation has been not reported so far.\\
We present some results collected during the first year of \INTEGRAL guaranteed time observations.
This paper is structured as follows: in section 2 we describe 
the data set we have used and the performed analysis.
In section 3, we report on imaging, temporal, and spectral analysis results 
obtained with the imager IBIS. We focus on the temporal behaviour
and spectral variability of a data sub-set, that is when the instrument 
achieves its best performances.

\begin{figure}[!t]
\centering
\includegraphics[width=7.9cm,height=7.6cm]{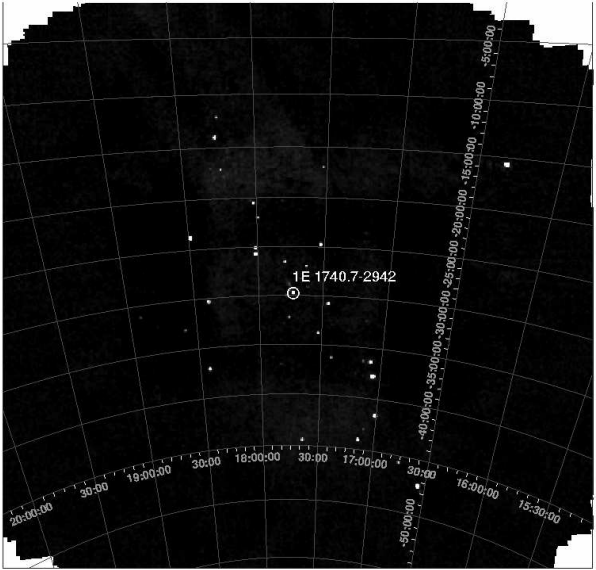}
\caption{20-40~keV significance maps mosaic of all GCDE SCWs (370) those include 1E 1740.7-2942 
(in the centre of the map), spanning from rev. 100 to rev. 122.}
\label{fig:pc_20_40}
\end{figure}

\section{Observations and Data Analysis}
\subsection{The \INTEGRAL monitoring}
The GC region has been frequently pointed by the \INTEGRAL satellite \cite{winkler03} 
during its guaranteed time observations (Core Programme).
The Core Programme consists of a deep exposure of the Galactic Centre (GCDE) and regular 
scans of the Galactic plane (GPS), plus some pointed observations.
During the first year the GCDE observing time was 4.8~Ms; as part of these observations,
the region including \mbox{1E 1740.7-2942} has been pointed for $\sim$1.4~Ms.\\
In this paper we present results of the first year of monitoring obtained with 
IBIS \cite{ubertini03}, the imager on board \INTEGRAL satellite.\\
IBIS is a coded mask telescope consisting of two detector layers, 
ISGRI \cite{lebrun03} and PICsIT \cite{labanti03};
herein we refer only to the data collected by the ISGRI detector.
The IBIS Partially Coded Field Of View (PCFOV) is 29$^\circ \times$29$^\circ$ 
at zero response, but the full instrument sensitivity 
is achieved in the 9$^\circ \times$9$^\circ$ Fully Coded Field of View (FCFOV).\\
By considering the coding effect of the instrument, during the first year of Core Programme 
the effective exposure time for \mbox{1E 1740.7-2942} was $\sim$790~ks. 
Basically, such a value (corrected for the dead-time) is calculated by multiplying 
the pointing exposure time by a vignetting map which is equal to 1 in
the FCFOV and decreases to zero at the borders of the PCFOV.\\
The total number of pointings, or science windows (SCWs), is 782, 
covering a period of 7 months between March 2003, revolution (rev.) 50,  
and October 2003, rev. 122 (the \INTEGRAL Julian Date interval corresponds 
to 1166.39 $\div$ 1382.77).\\
Raw data have been pre-processed and distributed by the Integral Science 
Data Center \cite{courvoisier03}.  
The IBIS scientific data analysis has been performed by using the \INTEGRAL 
off-line analysis software (OSA) release 3.0 \cite{goldwurm03}, 
Ftools 5.2 adapted to the \INTEGRAL fits files format by the HEASARC 
software team \footnote{\texttt{http://isdcul3.unige.ch/\~{}ebisawa/ftools.html}}.
The energy ranges selected for the imaging and temporal analysis are: 
20-40~keV, 40-80~keV, 80-120~keV, and 120-250~keV.
The light curves have been extracted from the images of each SCW: 
they have a binning time of $\sim$1800~s that is the mean duration of each GCDE pointing.\\ 
A logarithmic binning table with 16 energy channels has been chosen for spectral extraction.
Note that the response matrix for partially coded sources is not available yet, 
therefore we performed the spectral analysis of a sub-set of the whole data set: 
66 pointings with the source in the 
FCFOV for a total exposure time of $\sim$120~ks.\\
We used a {\it{renormalized}} effective area, in the sense that some 
systematics may be partially taken into account by introducing {\it{post facto}}
corrections using the Crab as a standard candle \cite{ubertini04};
in spite of this, we still need to assume a 2\% systematic error in order to obtain good fits.\\ 
Spectral fitting have been performed with the standard XSPEC v.11.2 tools and
all parameters errors have been calculated at 90\% of confidence level.

\begin{figure}[!t]
\centering
\includegraphics[width=8cm,height=7cm]{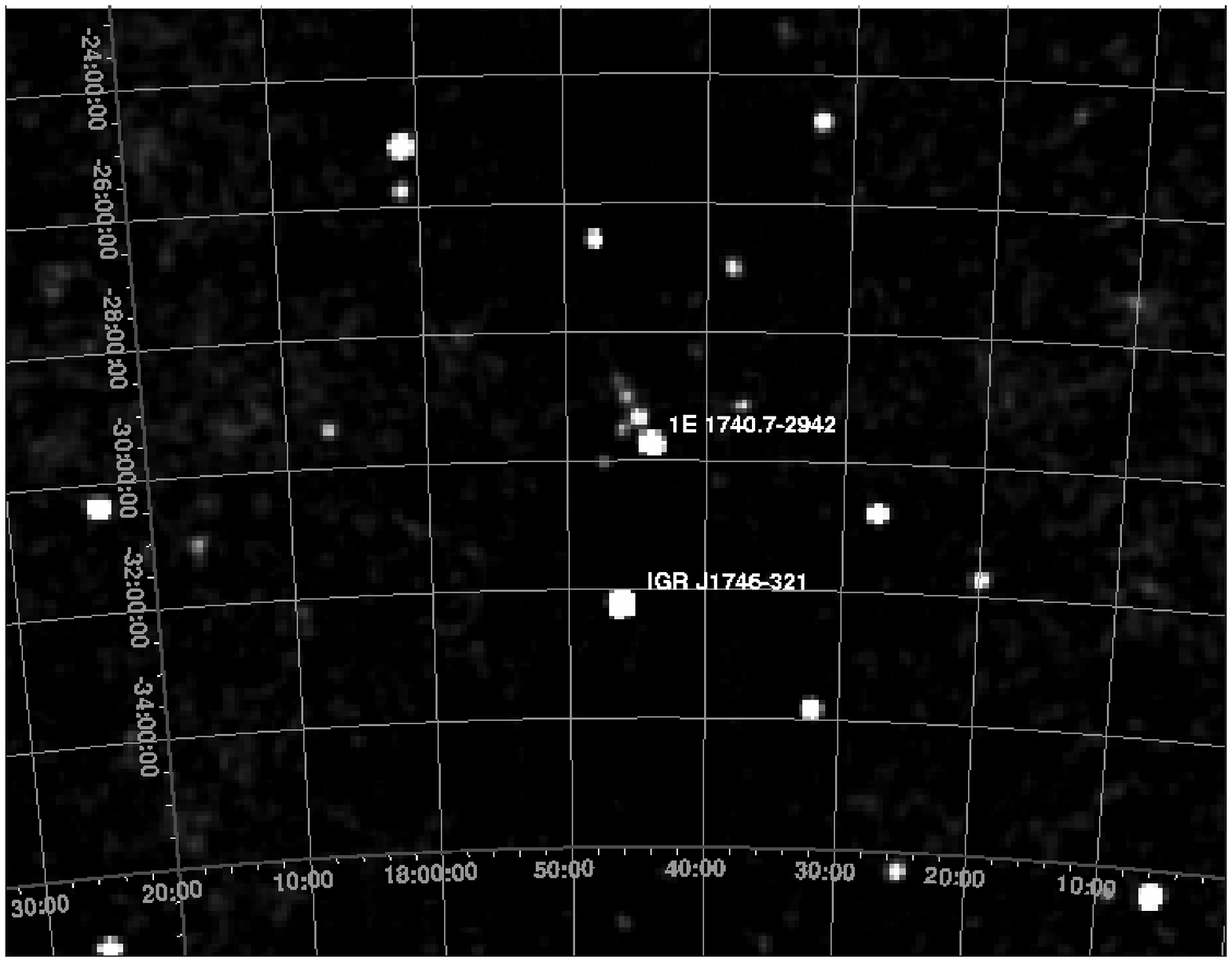}
\includegraphics[width=8cm,height=7cm]{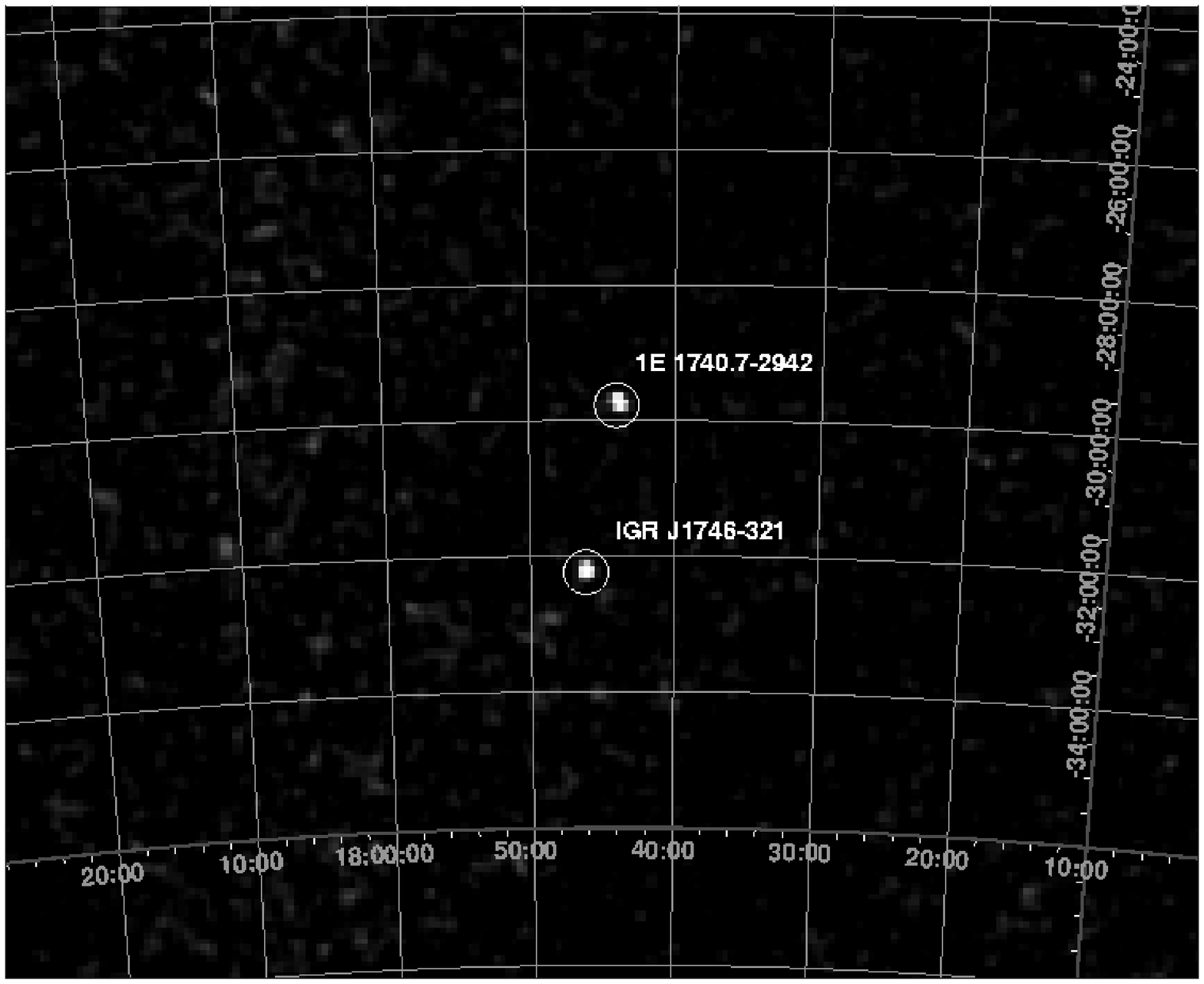}
\caption{Significance maps in the 20-40~keV (left) and 120-250~keV (right) bands of 
66 pointings when \mbox{1E 1740.7-2942} was in the IBIS FCFOV.}
\label{fig:fv_20_40}
\end{figure}

\section{Results}

\subsection{The {\it {RXTE~}} campaign}
The Rossi X-ray Timing Explorer ({\it{RXTE}}) 
Proportional Counter Array (PCA) has been periodically
monitoring \mbox{1E 1740.7-2942} since the first week of spacecraft
science operations in 1996.\\
These began as monthly pointings
of 1000-1500~s and have increased over the years until they
take place twice a week. This frequency has been found to
sample some of the more rapid state changes, and the
data quality is sufficient to look for small spectral changes
within the standard so called low/hard state, and to measure the
approximate level of fast variability, for each pointing.\\
Nearby sources are excluded by placing the centre of the 1$^\circ$
field of view about 0.5$^\circ$ away from \mbox{1E 1740.7-2942},
and the strong Galactic diffuse background is subtracted
with the help of pointings to an empty field symmetrically
positioned across the GC.\\  
This monitoring
campaign has been used to discover an unexpected temporal relation
between spectral shape and luminosity in this source, very
different from Cyg X-1 \cite{smith_a}; to identify
state changes promptly \cite{smith_b}; and to find the
orbital period of the \mbox{1E 1740.7-2942} system, 
estimated as 12.73$\pm$0.05 days \cite{smith_c}.

\begin{figure}[!t]
\centering
\includegraphics[width=4.5cm,height=7.8cm,angle=90]{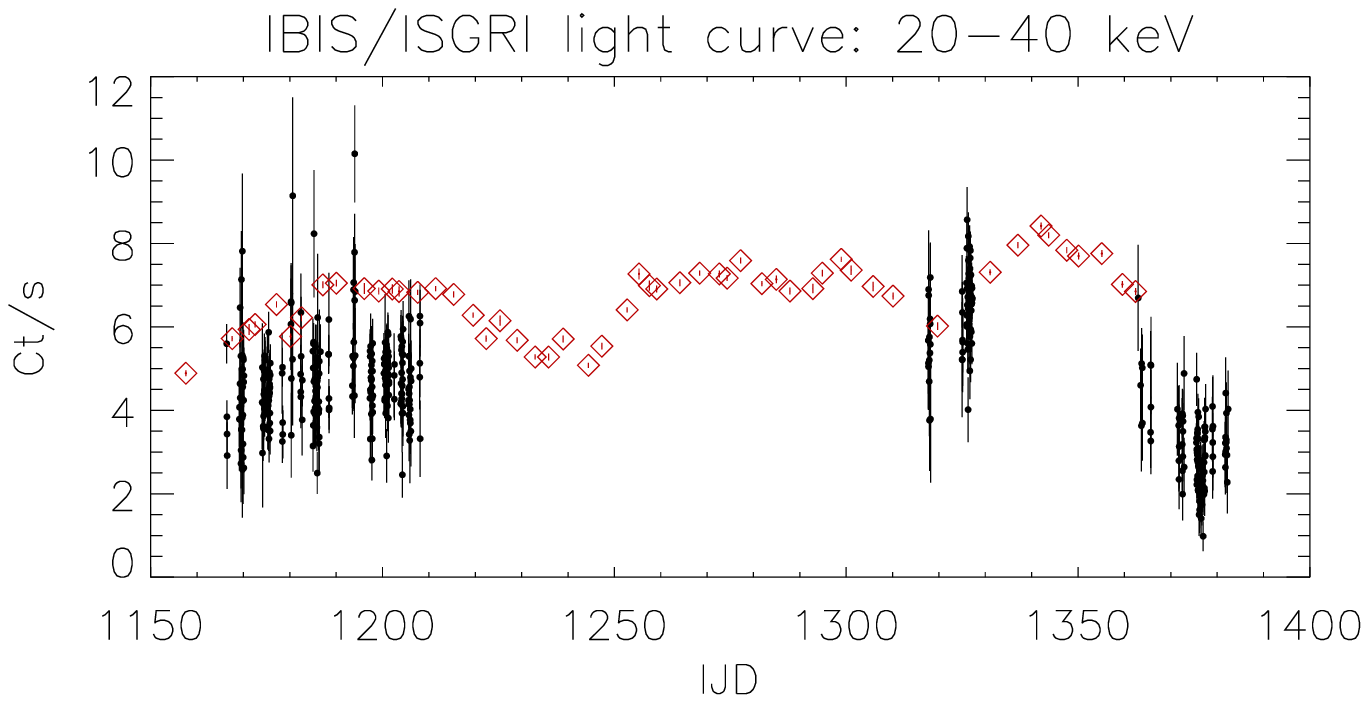}
\includegraphics[width=4.5cm,height=7.8cm,angle=90]{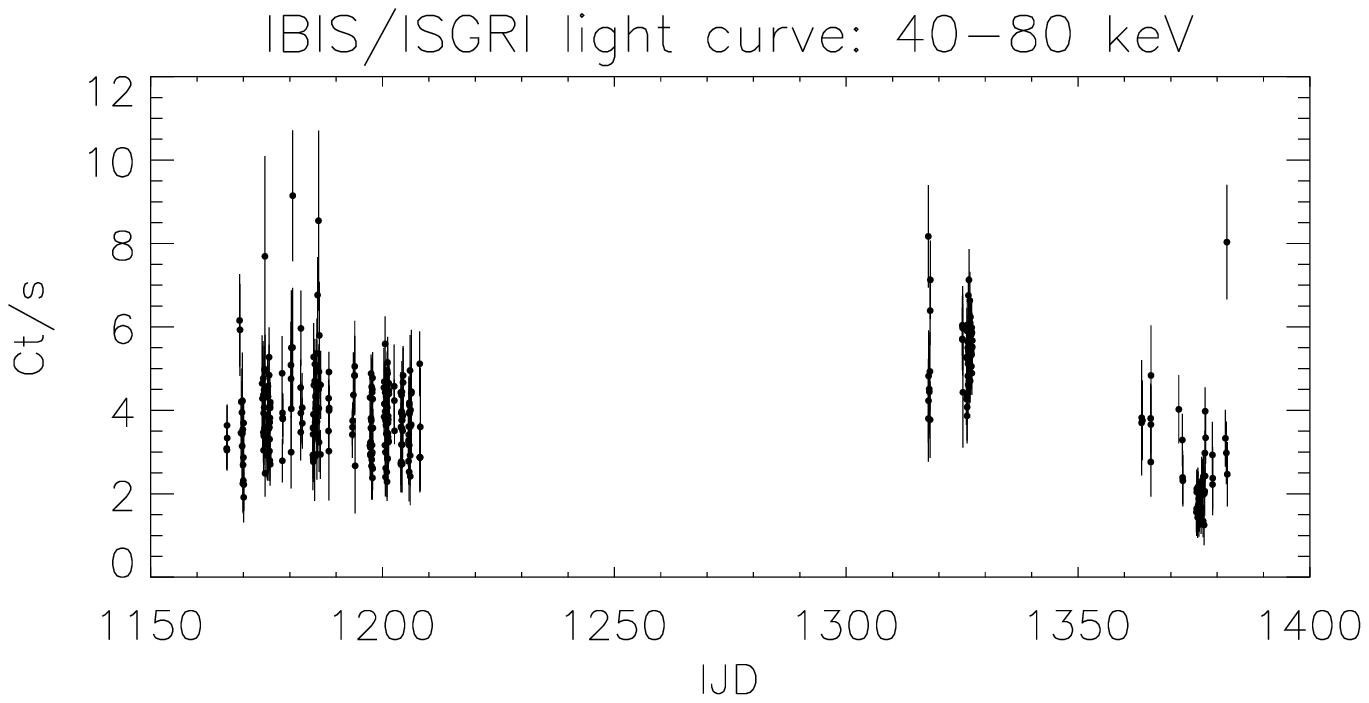}
\includegraphics[width=4.5cm,height=7.8cm,angle=90]{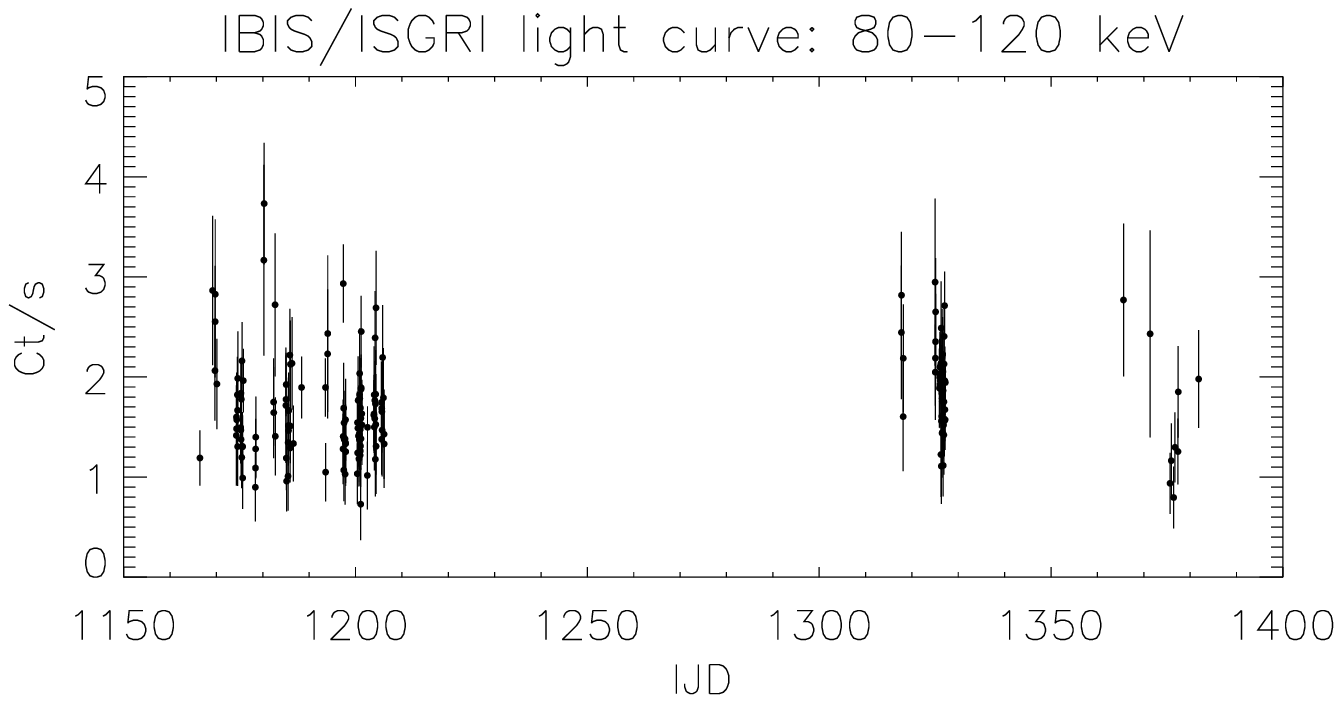}
\caption{Light curves extracted from the whole data set for 3 ISGRI bands. 
On top, the rhomboidal points superimposed are the RXTE/PCA data per detector from 8-25~keV.}
\label{fig:lc_t}
\end{figure} 

\subsection{Imaging analysis}
In the FCFOV of individual maps (corresponding to a single SCW), 
\mbox{1E 1740.7-2942} is detected in the range 20-40~keV  
at a level of confidence of $10 \sigma$  and
at a level of $3 \sigma$ at $14^\circ$ from the centre of the FOV.\\ 
In the energy range 40-80 keV, it is detectable up to offset angles of 
$\sim$12$^\circ$; on the contrary, between 80 and 120 keV it has 
been detected in only a few pointings (less than 200 SCWs).
The 20-40~keV mosaic of 370 significance maps ($6.15\times 10^5$~s) 
is shown in Figure \ref{fig:pc_20_40}: 
numerous sources are detected by IBIS/ISGRI around
\mbox{1E 1740.7-2942}, which is in the centre of 
the map at a detection level of $112 \sigma$.\\  
In Figure \ref{fig:fv_20_40} two mosaics for two different energy bands built with 66 maps
when \mbox{1E 1740.7-2942} was in the FCFOV are shown.
In the GC zone, the only two sources detected up to 250~keV 
are two BHCs: \mbox{1E 1740.7-2942} and the transient source 
\mbox{IGR J1746-3213} \cite{capitanio04}.

\begin{figure*}
\centering
\begin{tabular}{cc}
\includegraphics[width=4.5cm,height=8cm,angle=90]{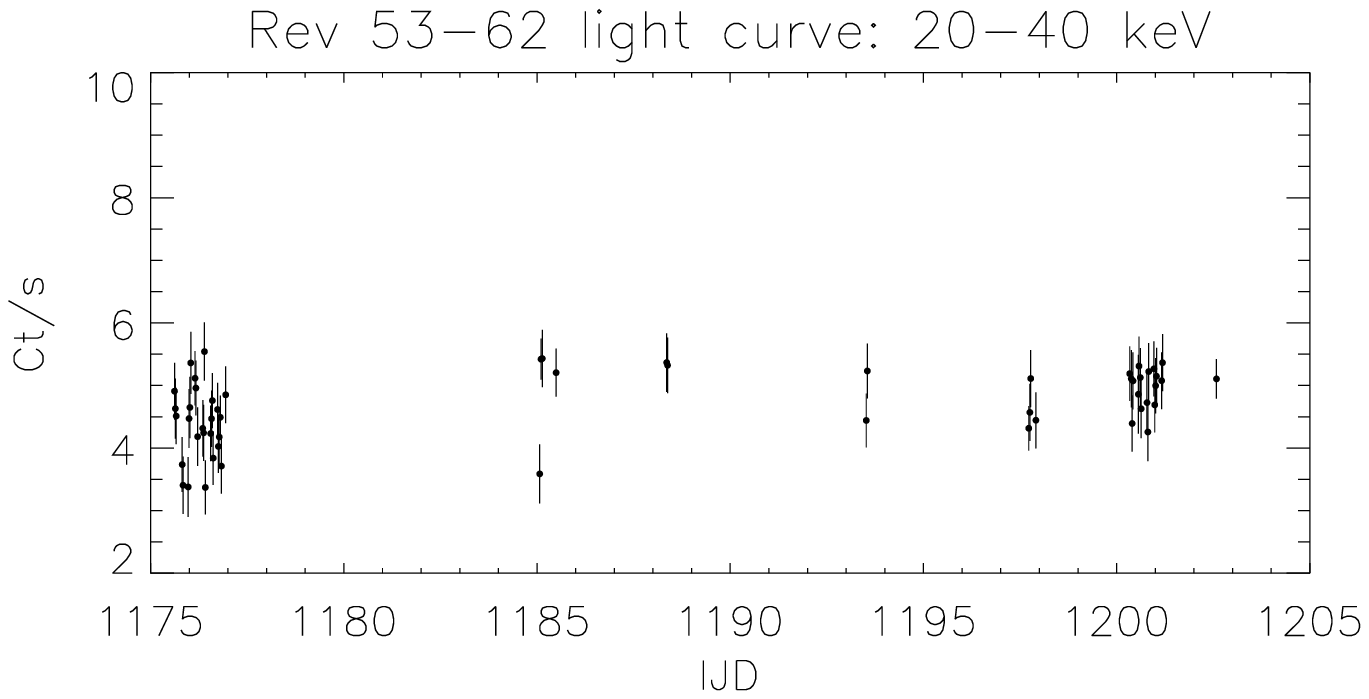}&
\includegraphics[width=4.5cm,height=8cm,angle=90]{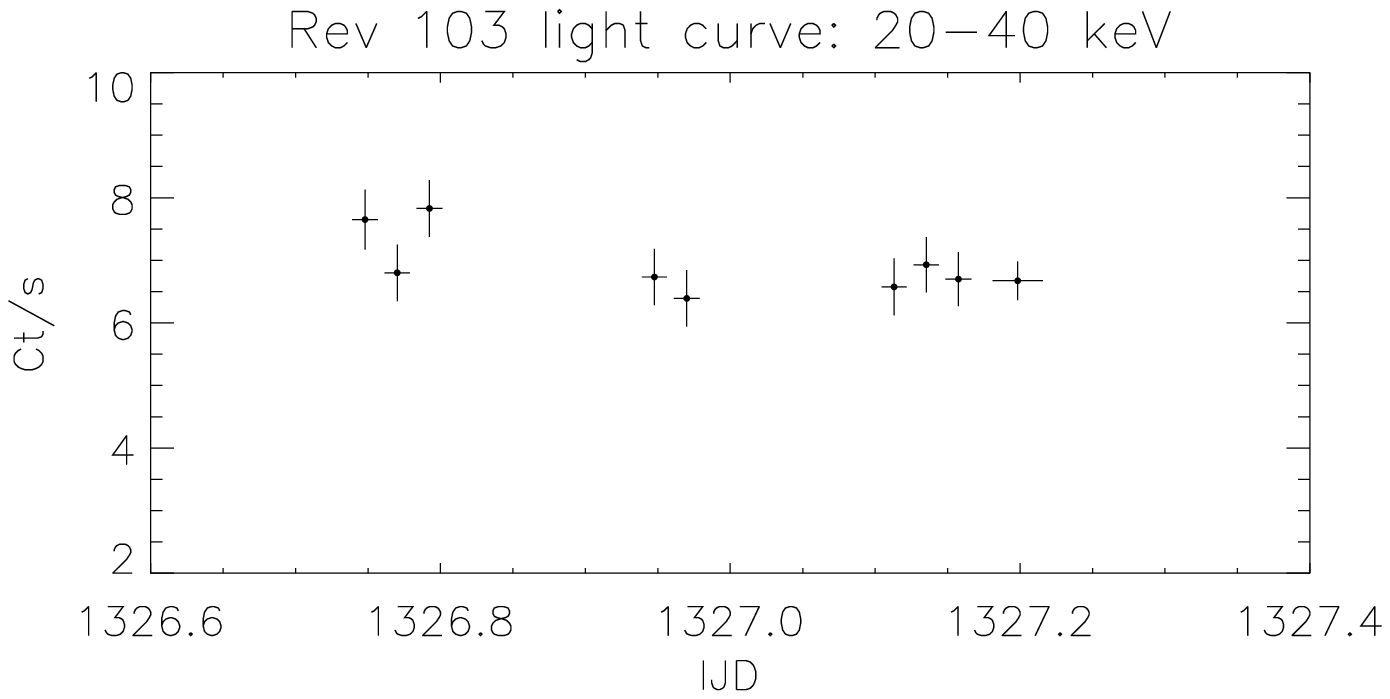}\\
\includegraphics[width=4.5cm,height=8cm,angle=90]{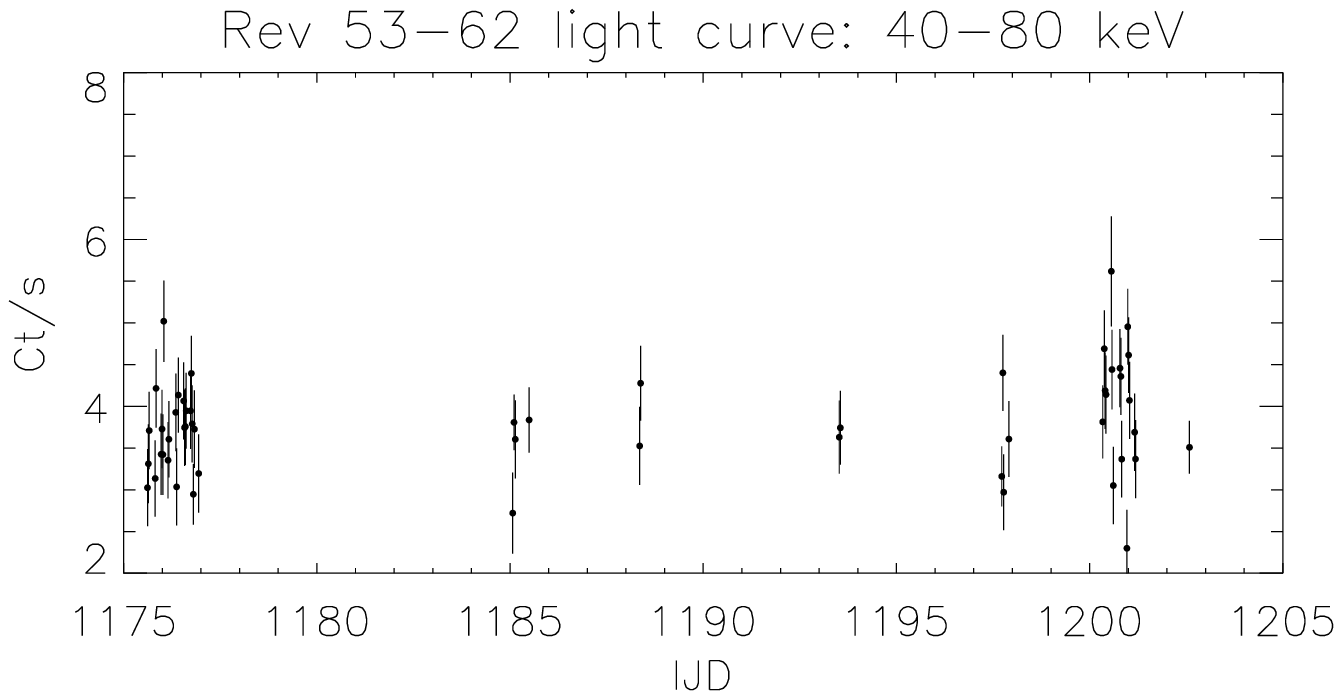}&
\includegraphics[width=4.5cm,height=8cm,angle=90]{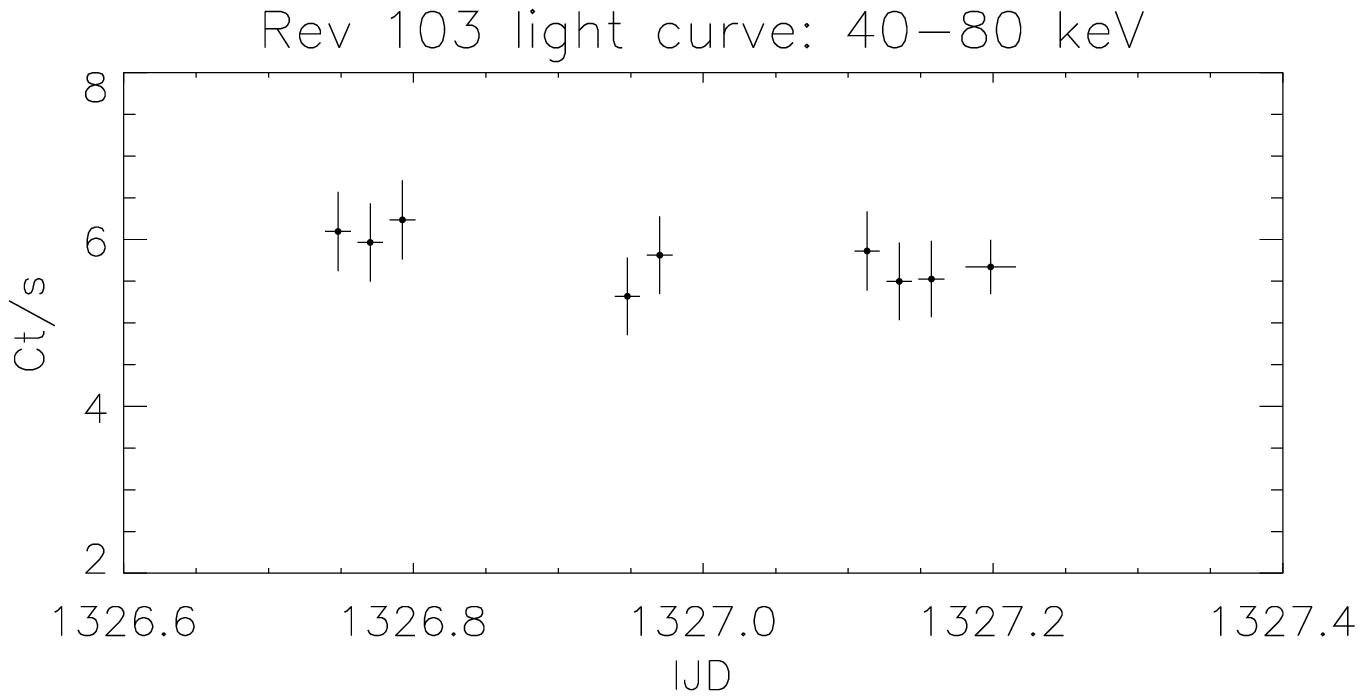}\\
\end{tabular}
\caption{Light curves of the FCFOV observations in two different energy bands:
the flux clearly increased comparing the first (rev. 53-62, left) 
and the second (right) part of the observation.}
\label{fig:lc_f}
\end{figure*}

\subsection{Light curves}

Light curves corresponding to the first 3 selected energy ranges are shown in Figure \ref{fig:lc_t}. 
At low energy we detected \mbox{1E 1740.7-2942} in 460 SCWs.
We decided to reject all the SCWs with statistical errors not determined by the software.\\
On the first plot (top) the {\it{RXTE}}/PCA light curve in the energy range 8-25~keV is superimposed. 
As can be seen, the two temporal profiles look consistent during the simultaneous observations.\\
In the two energy ranges 40-80~keV (middle) and 80-120~keV (bottom), 
the number of good SCWs we found is 371 and 180 respectively.\\
Some systematic errors could still affect the count rate evaluation, 
because the instrumental off-axis response has not been fully evaluated yet.
Because of that, we also extracted light curves considering only SCWs with the source in the FCFOV. 
These observations cover three time intervals: rev. 53-62, rev. 103, and rev. 119-120.\\
As it is shown in Figure \ref{fig:lc_f}, there is no evidence of flux 
variation within each of the two periods, 
while on a longer time scale there is a clear indication of a 30\% increase.
The mean value of the count rate and the variance have been calculated in the energy range 20-40~keV:
with OSA 3 software we obtain 4.7$\pm$0.3 ct/s and 6.9$\pm$0.2 ct/s 
as mean rates within the first and second period respectively.\\ 
Considering mCrab, such increase comes from 41~mCrab up to 
61~mCrab\footnote{The Crab gives 114~ct~s$^{-1}$ in ISGRI from 20-40~keV.}.
On the contrary, during the third period the source decreased above 30 mCrab.
The later (more recent) INTEGRAL observations \cite{atel} showed that the source flux 
declined below the ISGRI sensitivity limit ($\sim$1.7 mCrab, 3 $\sigma$) by March 2004.\\
Hereafter we refer to the spectral analysis performed by using the first and the 
second period data set.\\

\subsection{Hardness ratio and spectral behaviour}
\begin{figure}[!t]
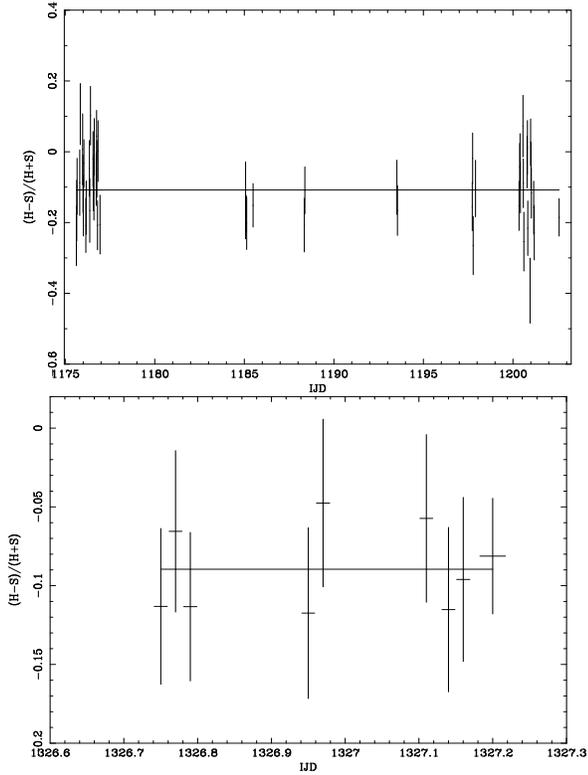

\centering
\includegraphics[width=5.2cm,height=7.3cm,angle=-90]{qdp_1st_ref.ps}
\includegraphics[width=5cm,height=7.7cm,angle=-90]{qdp_2nd_ref.ps}
\caption{Hardness ratios from the first period observations (rev 53-62, top) 
and the second one (rev 103, bottom); the continuous line is the best fit.}
\label{fig:hr}
\end{figure}

\begin{figure}[!b]
\centering
\includegraphics[width=5cm,height=8cm,angle=+90]{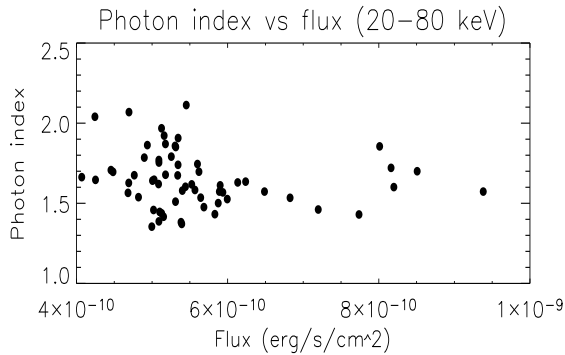}
\caption{Photon index versus flux estimated SCW by SCW; the photon index errors are roughly 20\%.}
\label{fig:fl_ind}
\end{figure}

\begin{figure}[!t]
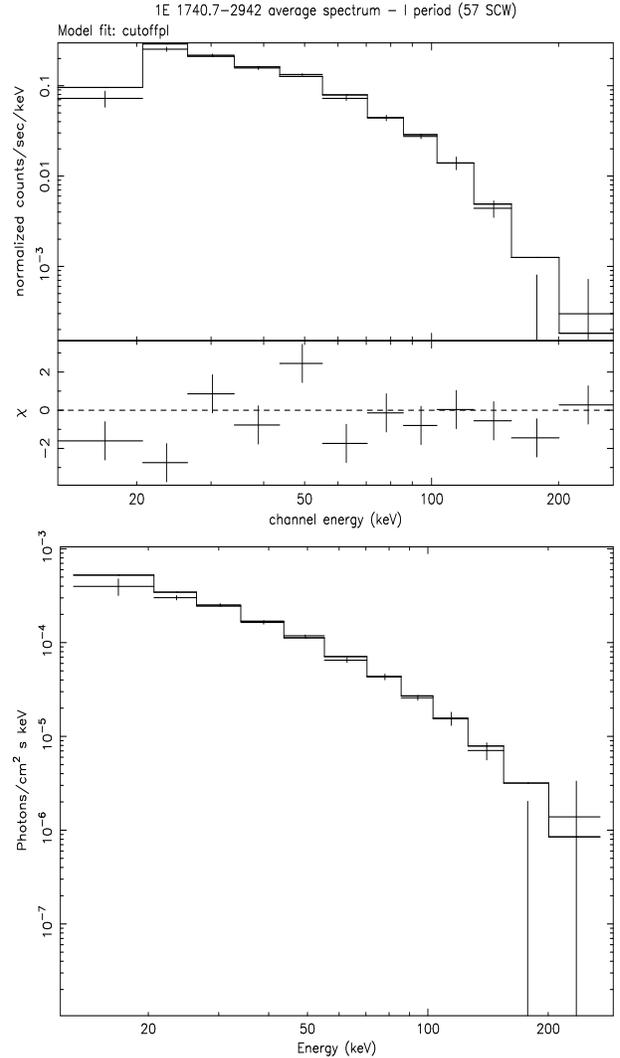

%\centering
%\begin{tabular}{cc}
\centering
\includegraphics[width=7cm,height=8cm,angle=-90]{ct_spec_I.ps}
\includegraphics[width=7cm,height=8cm,angle=-90]{ph_spec_I.ps}
%\end{tabular}
\caption{Averaged spectrum of the \mbox{1E 1740.7-2942} during the first period 
fitted by a power law with high energy cut-off.}
\label{fig:cpl_I_P}
\end{figure}

\begin{figure}[!t]
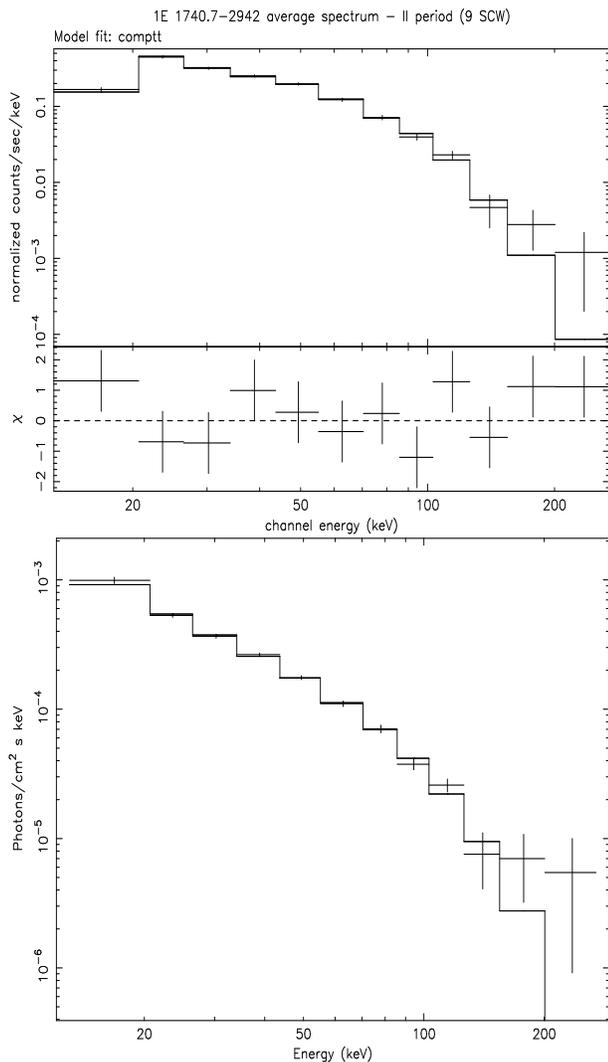

%\centering
%\begin{tabular}{cc}
\centering
\includegraphics[width=7cm,height=8cm,angle=-90]{ct_spec_II.ps}
\includegraphics[width=7cm,height=8cm,angle=-90]{ph_spec_II.ps}
%\end{tabular}
\caption{Averaged spectrum during the period with higher flux fitted with the \texttt{compTT} model.} 
\label{fig:comptt}
\end{figure}

In order to search for spectral variations, we evaluated the hardness ratio defined as: 
\begin{eqnarray}
HR & = & \frac{(h-s)}{(h+s)}
\end{eqnarray} 
with $h$ and $s$ being the count rate in the 40-80~keV and 20-40~keV energy ranges respectively.\\
Performing a $\chi^2$ test, no spectral variations have been found within each of the two periods.
In Figure \ref{fig:hr}, the hardness ratios and the models employed for the fit have been plotted: 
both fits are consistent with a constant model with roughly the same value.\\  
In addition, all spectra (in the range 20-80~keV) extracted from 
each SCW have been fitted by a single power law.
In a single GCDE pointing, the source intensity and low counting statistics 
do not allow us to constrain the cut-off. 
The aim of this analysis was to investigate a possible correlation
between the photon index and the flux: 
no relation has been found, as is clearly visible in Figure \ref{fig:fl_ind}.\\  
Since a simple power law does not allow a good fit ($\chi_{\nu}^2/(9)$=7),
the averaged spectrum corresponding to the first period (57 spectra)  
has been fitted by a cut-off power law (Figure \ref{fig:cpl_I_P}).
A fit with a $\chi_{\nu}^2/(9)$=1.4 has been obtained with
a photon index $\Gamma=0.43_{-0.32}^{+0.19}$ and a high energy cut-off $E_{c}=36\pm6$~{keV}.\\ 
We also fitted the averaged spectrum from the second data set, consisting 
of 9 spectra at the higher flux level,
both using the empirical cut-off power law model and the physical thermal Comptonization \cite{tita} 
model (\texttt{compTT} in XSPEC).
The empirical model gives $\Gamma=0.84_{-0.24}^{+0.22}$ 
and $E_{c}=36_{-10}^{+14}$~keV with a $\chi_{\nu}^2/(9) = 1.1$.\\
In the two fluxes state, even though the parameters are a little bit different, 
they are consistent within the errors, confirming the hardness ratio result.\\
A good fit ($\chi_{\nu}^2/(9)=1.2$) on the second data set has been also obtained 
by using the \texttt{compTT} model (Figure \ref{fig:comptt}).
Following, the Comptonization parameters that we have found:
a plasma temperature (kT) of $22\pm2$~keV and a Thompson optical depth $\tau=2.5\pm0.3$.
The flux estimated between 20 and 250~keV was $1.2 \times 10^{-9}$~erg~cm$^{-2}$~s$^{-1}$.\\
The \texttt{compTT} model does not allow a good fit to the first data set because the low statistic
does not permit to constrain the fit parameters. 

\section{Conclusion}
The BHC \mbox{1E 1740.7-2942} has been detected by IBIS/ISGRI up to 250 keV in $\sim$120~ks.
During the seven months of \INTEGRAL monitoring, 
we measured for this source a flux variation up to 50\% on a time scale of months, but
without any change in spectral shape. The source 20-40 keV flux
increased from 40~mCrab up to 60~mCrab and then decreased 
to roughly 30~mCrab during October 2003.\\ 
In the IBIS/ISGRI energy range, the \mbox{1E 1740.7-2942} 
spectral shape is typical of the BHCs in their low/hard state.\\
The empirical model achieves an energy cut-off of about 40~keV, while  
a 22~keV Comptonization temperature and an optical depth $\tau=2.5\pm0.3$ have been found
by using the thermal Comptonization fit.
Such a result is consistent with the BeppoSAX Galactic Centre observations performed 
with the source at the same flux level \cite{sidoli99}.\\
In order to study the source behaviour by using a broad band spectrum, we plan to 
continue with analysis of the JEM-X and SPI data as well.\\
In conclusion, we underline the IBIS fine imaging capabilities,
especially in crowded field like the Galactic Centre.

\section*{Acknowledgments}
This paper is based on observations with INTEGRAL, an ESA project
with instruments and science data centre funded by ESA member states (especially the PI countries:
Denmark, France, Germany, Italy, Switzerland, Spain), Czech Republic and Poland, 
and with participation of Russia and USA.\\
MDS, FC, GDC, AM, AP acknowledge financial support from the Italian Space Agency (ASI).\\
The work described in this paper has been partially supported by ASI.\\
The IBIS team at IASF-Rome thanks Memmo Federici for the hardware support and data archiving.\\
MDS thanks Alessandra De Rosa for precious suggestions.\\

%\begin{small}
%\begin{verbatim}

%\end{verbatim}
%\end{small}

\end{document}